\documentclass[a4paper,11pt]{article}
\pdfoutput=1 

\usepackage{jheppub}  

\usepackage[T1]{fontenc}  

\usepackage{amsfonts}

\title{Chern-Simons branes with enhanced gauge invariance}
 
\author[a,1]{Pablo Mora,\note{Corresponding author.}}
 
\affiliation[a]{Centro Universitario Regional Este (CURE), Universidad de la Rep\'ublica, Uruguay, Ruta 9 km. 207, Rocha, Uruguay}
 
\emailAdd{pablomora@cure.edu.uy}
 
\abstract{I discuss how the factorization of the invariant trace used to define Chern-Simons branes in a space-time
with a Chern-Simons action for a space-time group introduces new relationships between 
the coupling constants of the extended objects of diverse dimensions, and an enhanced gauge invariance for a suitable choice of these
coupling constants, owing to an extension of the Inflow mechanism. 
I also comment on the possible relevance of these models in fundamental physics.}

\begin{document} 
\maketitle
\flushbottom

\section{Introduction}

Chern-Simons branes where introduced in Ref.\cite{Mora-Nishino20001} as 
extended objects in Chern-Simons (super)gravity theories \cite{Achucarro:1987vz,Chamseddine:1990gk, Chamseddine:1989nu, MuellerHoissen:1990vf, Banados:1993ur, Banados:1996hi, Zanelli:2012px}. The motivations and lines of research weaved in these models were discussed in the Introduction to my previous Letter \cite{Mora:2021wmz}, with the relevant 
References \cite{Dixon:1991xz,Green:1989nn,Troncoso1997va, Troncoso1998ng,Horava:1997dd}. Also references to other works on Chern-Simons branes were given \cite{Edelstein:2008ry,Miskovic:2009dd,Edelstein:2010sx,Edelstein:2010sh,Edelstein:2011vu,Kastikainen:2020auf,Frey:2019fqz,Ertem:2012qv}.\\
The ultimate goal of this line of research is to formulate a coherent and simple framework founded on deep physical principles.
Wheeler once wrote \cite{Wheeler} "Physics must be in the end law without
law. Its undergirding must be a principle of organization which is no organization at all. In all of mathematics,
nothing of this kind more obviously offers itself than the principle that the boundary of a boundary is zero. That this
principle should pervade physics, as it does -- is that the only way that nature has to signal to us a construction
without a plan, a blueprint for physics that is the very epitome of austerity?". Both Chern-Simons gravities and 
Chern-Simons branes models embody this principle, but in the opposite form, "the boundary of the boundary is not zero ", 
as shown for instance in the descent equations in Section 2.\\ 
Wheeler's quote \cite{Wheeler} continues: "No. A second sign directs the seeker for the plan of existence still more clearly to austerity: the quantum.
What is the thread that connects mystery number two, the quantum, with puzzle number one, the machinery of
existence?". Quantum Mechanics fundamental notions \cite{Dirac} are quantum states and phases, and the transformation theory of unitary operators\footnote{Observables are given by the phases associated to these transformations, for instance Energy to time displacements, 
Momentum to spacial translations and Angular momentum to rotations (see for instance first footnote in Ch.VI in Ref.\cite{Weinberg}).} (that are themselves non abelian phases).
These phases are directly connected to the Gauge Principle, as Yang \cite{Yang} has remarked. Chern-Simons gravities \cite{Achucarro:1987vz,Chamseddine:1990gk, Chamseddine:1989nu, MuellerHoissen:1990vf, Banados:1993ur, Banados:1996hi, Zanelli:2012px} are the answer to the question of how to write a theory including gravity as a gauge system, and Chern-Simons branes models \cite{Mora-Nishino20001,Mora:2021wmz} allow to introduce extended objects in these theories in a gauge invariant and background independent way\footnote{
This differs from String Theory, where gauge invariance appears in an indirect way, except in the case of the 
Heterotic String \cite{Gross:1984dd,Gross:1985fr,Gross:1985rr}, and in the appearance in String Theory of a space-time metric and p-form fields. In Chern-Simons models the metric is built from certain components of the gauge field identified as a vielbein and the p-form fields are composite fields given by Chern-Simons forms of the suitable order, which automatically yields the anomalous gauge transformations required.}.\\
It was also suggested in Refs.\cite{Mora-Nishino20001,Mora:2021wmz} that the structure of the action as a sum of differential forms
of similar characteristics integrated in manifolds of the corresponding dimension corresponds to giving physical content to the differential structure of space-time, as General Relativity gives physical content to its metric structure.\\
In this article I continue the study of Chern-Simons branes, particularly the consequences of the factorization, when it occurs,
of the invariant trace used to define these theories. These consequences are new relationships between 
the coupling constants of extended objects of diverse dimensions and the space-time, and an enhanced gauge invariance under variations of just one of the two gauge fields required to define the Transgression forms, for a suitable choice of the coupling constants.
These results are important as they allow to narrow the set of possible theories.\\
The plan of this work is as follows: In Section 2 I review the Green-Schwarz and Inflow anomaly cancellation mechanisms. This review 
section is necessary because I will reformulate these results in the context of Chern-Simons branes. 
Section 3 reviews the essentials of Chern-Simons branes, which are the subject of the rest of the article. Section 4 deals with the relationships between coupling constants that 
occur if the trace factorizes, and Section 5 introduces "electric" and "magnetic" charges in that case. Section 6 discussed the above mentioned enhanced gauge invariance. Section 7 provides concrete examples of trace factorization.
Finally in Section 8, Discussion and Conclusions, I comment on the present work and outline future lines of research. 

\section{Review of the Green-Schwarz and Inflow anomaly cancellation mechanisms}

\subsection{Anomalies}

The gauge invariant anomaly polynomial\footnote{For the topic of Anomalies see 
Refs.\cite{Zumino:1983rz,Alvarez-Gaume:1983ihn,Alvarez-Gaume:1984zlq} and References therein.} 
$\Omega _{2n+2}$ is a $(2n+2)$-form is related to the Chern-Simons $(2n+1)$-form by
$\Omega _{2n+1}$ by
\begin{equation*}
\Omega _{2n+2}=d\Omega _{2n+1}.
\end{equation*}
It can also be written as an invariant trace $<something>$ of the exterior product of $n+1$ curvature (field strength) 2-forms $F$ as
\begin{equation*}
\Omega _{2n+2}=<F^{n+1}>.
\end{equation*}
The gauge invariance of the anomaly polynomial $\delta \Omega _{2n+2}=0$ results from the fact that for the invariant trace 
$d<something>= <D(something)>$, where $D$ is the covariant exterior derivative, and the Bianchi identity $DF=0$. It implies that 
the gauge variation of the Chern-Simons form must be closed, and hence locally exact $\delta\Omega _{2n+1}=d\Omega ^1_{2n}$. 
The   Wess-Zumino-Witten (WZW) $2n$-form $\Omega ^1_{2n}$ gives the anomaly 
as the variation to the quantum effective action $\delta\Gamma =\int \Omega ^1_{2n}$.

\subsection{The Green-Schwarz mechanism}

The anomaly can be cancelled by the Green-Schwarz mechanism \cite{Green:1984sg} if the anomaly polynomial 
factorizes as the exterior product of two anomaly polynomials\footnote{Through all this work we use a notation where the exterior (wedge) product of differential forms is implied 
$\alpha _p\wedge \beta _q\equiv \alpha _p\beta_q$.} $\Omega _{2n+2}=\Omega _{p+2}\Omega _{2n-p}$, with 
$d\Omega _{p+2}=0$, $\delta \Omega _{p+2}=0$, $d\Omega _{2n-p}=0$ and  $\delta \Omega _{2n-p}=0$. 
Also $\Omega _{p+2}=d\Omega _{p+1}$, $\Omega _{2n-p}=d\Omega _{2n-p-1}$, where $\Omega _{p+1}$ and $\Omega _{2n-p-1}$ are
the corresponding Chern-Simons forms. As before, the Chern-Simons and corresponding WZW forms satisfy  that locally
$\delta\Omega _{p+1}=d\Omega ^1 _{p}$ and $\delta\Omega _{2n-p-1}=d\Omega ^1 _{2n-p-2}$. 
It follows that 
$\Omega _{2n+2}=\Omega _{p+2}\Omega _{2n-p}=d\Omega _{p+1}\Omega _{2n-p}=d[\Omega _{p+1}\Omega _{2n-p}]$, 
because $d\Omega _{2n-p}=0$.
Alternatively $\Omega _{2n+2}=\Omega _{p+2}\Omega _{2n-p}=\Omega _{p+2}d\Omega _{2n-p-1}=d[\Omega _{p+2}\Omega _{2n-p-1}]$.
The anomaly is then given by
\begin{equation*}
\delta\Gamma = \int \Omega _p^1\Omega _{2n-p}=\int \Omega _{p+2}\Omega _{2n-p-2}^1,
\end{equation*}
as $d[\Omega _p^1\Omega _{2n-p}]=\Omega _{p+1}\Omega _{2n-p}$, 
or $d[\Omega _{p+2}\Omega _{2n-p-2}^1]=\Omega _{p+2}\Omega _{2n-p-1}$. The Green-Schwarz mechanism states that if the theory includes 
a p-form field $B_p$ with an anomalous gauge transformation $\delta B_p=-\Omega _p^1$ 
and an additional term in the Lagrangian of the form $\Delta L=B_p\Omega _{2n-p}$, so that the action goes from
$\Gamma$ into
\begin{equation*}
\Gamma \rightarrow \Gamma +\int \Delta L=\Gamma +\int B _p\Omega _{2n-p},
\end{equation*}
then the new action is anomaly free, that is
\begin{equation*}
\delta [\Gamma +\int \Delta L ]=0.
\end{equation*}
Notice that the gauge invariant field strength is not $dB_p$, but 
\begin{equation*}
H_{p+1}=dB_p+\Omega _{p+1}.
\end{equation*}
In the case of String Theory the field $B_p$ and its anomalous gauge transformation do arise from a perturbative expansion,
with the anomalous transformation a one-loop order contribution.

\subsection{Inflow mechanism}

The Inflow mechanism was introduced by Callan and Harvey
\cite{Callan:1984sa}, and 
was discussed in Duff, Liu and Minasian \cite{Duff:1995wd} and by 
Witten \cite{Witten:1995em} in the case of 
five-branes in an eleven dimensional background space-time. This mechanism is related to, 
but different from, the Green-Schwarz mechanism, and 
it also requires factorization of the invariant trace. 
Other relevant works on this matter are Refs.\cite{Dixon:1992if,Duff:1994vv,Vafa:1995fj,Witten:1996md}.

Following Ref.\cite{Witten:1995em}, we consider a 5-brane 
in an eleven dimensional space-time described by supergravity \cite{Cremmer:1978km}, with action
$\Gamma =\Gamma _{11D-SUGRA}+\Gamma _{5-brane}$.The eleven dimensional theory 
contains a 3-form potential $A_3$ and a 4-form field strength $F_4=dA_3$. Normally 
$F_4$ should be closed, that is $dF_4=0$, yet in presence of a 5-brane it satisfies  $dF_4=\delta _V$, 
where $\delta _V$ is a delta function
in the 5-brane world-volume. For instance, in one dimension and if $P$ is the origin, 
$\delta _P=\delta (x)dx$, satisfying $\int\delta _P=1$ and $d\delta _P=0$. This is not different from 
a fundamental point-like monopole solution in four dimensions, where $\int _M F_2 =g$, and $M$ is a two dimensional manifold enclosing the monopole with magnetic charge $g$, therefore instead of the standard Bianchi identity $dF_2=0$ we obtain $dF_2=g\delta _V$, with $V$ a one dimensional manifold corresponding to the world-line of the monopole, as the two dimensional manifold $M$ can contract indefinitely towards the monopole and the result is still $g$ \footnote{I am fully aware of the mathematical subtleties implied, related to the Dirac
string or  local charts. We use here and henceforth, as in Ref.\cite{Witten:1995em}, the delta function notation as a useful way to summarize the relevant relationships at the level of rigor of theoretical physics.} . 
Witten considers world-volume gravitational anomalies, resulting from
infinitesimal diffeomorphisms in the world-volume of the 5-brane under which the 5-brane action is not invariant, but rather transforms as 
$\delta \Gamma _{5-brane} =\int _V J_6$ where $J_6$ is the WZW term associated to that anomaly. In turn, $J_6$ is related to 
the gravitational Chern-Simons term $I_7$ as $\delta I_7=dJ_6$. Then Witten supposes that the eleven dimensional
action receives a correction $\Gamma_{11D-SUGRA}\rightarrow\Gamma _{11D-SUGRA}+\int _M F_4I_7$, and the new total action is
$\Gamma '=\Gamma +\int _MF_4I_7$. Under diffeomorphisms on the world-volume of the 5-brane 
$\delta \Gamma '=\delta \Gamma _{11D-SUGRA}+\delta\Gamma _{5-brane}+\delta\int _MF_4I_7=\int _VJ_6+\delta\int _MF_4 I_7$, as 
$\delta \Gamma _{11D-SUGRA}=0$. But 
$\delta\int _MF_4 I_7=\int _MF_4\delta I_7= \int _MF_4dJ_6  =-\int _M dF_4 J_6+boundary~term= -\int _M \delta _V J_6+boundary~term =-\int _V J_6+boundary~term$,
therefore the world-volume gravitational anomaly cancels for the modified action, that is $\delta \Gamma '=0$, assuming the 
space-time boundary term vanish.

\subsection{Quantization of constants for the five-brane}

In ref.\cite{Duff:1995wd} quantization conditions and relationships between the constants involved, namely 
the 2-brane and 5-brane tensions and the gravitational constant, were discussed, in the case of eleven dimensional 
supergravity coupled to branes. I review here these arguments, with an eye to a reformulation in the case of Chern-Simons branes.

Duff, Liu and Minasian \cite{Duff:1995wd} consider the bosonic part of the 2-brane action
\begin{eqnarray}
S_3=T_3\int d^2\xi[-\sqrt{-\gamma}\gamma ^{ij}\partial _iX^{M}\partial _jX^{N}g_{MN}(X)+\nonumber\\
+\frac{1}{2}\sqrt{-\gamma}-\frac{1}{3!}\epsilon ^{ijk}(A_3)_{MNP}\partial _iX^{M}\partial _jX^{N}\partial _kX^{P}]\nonumber
\end{eqnarray}
and the bosonic part of the eleven dimensional gravity action
\begin{equation*}
S_{11}=\frac{1}{2\kappa _{11}}\int d^{11}x\sqrt{-g}[R-\frac{1}{2.4!}(F_4)_{MNPQ}(F_4)^{MNPQ}]
-\frac{1}{12\kappa _{11}^2}\int A_3F_4F_4,
\end{equation*}
with $F_4=dA_3$. A Dirac-like quantization argument applied to the brane action shows that 
$T_3\int _{\mathcal{S}^3\equiv \partial\mathcal{S}^4}A_3=T_3\int _{\mathcal{S}^4}F_4=2\pi n$, 
with $n$ integer, or
$\int _{\mathcal{S}^4}F_4=\frac{2\pi n}{T_3}$. Then looking to the eleven dimensional action they consider 
$-\frac{1}{12\kappa _{11}^2}\int _{\mathcal{S}^{11}\equiv\partial \mathcal{S}^{12}} A_3F_4F_4=
-\frac{1}{12\kappa _{11}^2}\int _{\mathcal{S}^{12}} F_4F_4F_4$.
Using again a Dirac-like argument yields  
$-\frac{1}{12\kappa _{11}^2}\int _{\mathcal{S}^{12}} F_4F_4F_4=2\pi m$, 
with $m$ integer. While it is not true in general, but only in the case 
of product manifolds, that for differential forms $\alpha $ and $\beta$ it would be that
$\int \alpha\wedge\beta =\int\alpha\int\beta$, consistency of both results requires 
$3!.(\frac{2\pi }{T_3})^3n_1n_2n_3=2\pi m.12\kappa _{11}^2$, or $\frac{(2\pi)^2}{\kappa _{11}^2T_3^3}n_1n_2n_3=2m$,
which implies that $\frac{(2\pi)^2}{\kappa _{11}^2T_3^3}$ is an even integer.

To conclude this section, it is worthwhile to mention that, using anomaly cancellation and duality arguments, references \cite{Duff:1995wd, Witten:1995em} argue that the Chern-Simons-like part of the eleven dimensional Lagrangian
gets corrected from a term of the schematic form $A_3F_4F_4$ to $A_3F_4F_4+ A_3X_8$, with $X_8$ an anomaly polynomial given explicitly
in  \cite{Duff:1995wd}.

\section{Actions for Chern-Simons branes}

 As in Refs.\cite{Mora-Nishino20001,Mora:2021wmz}, I consider actions for extended objects (branes) with dynamical variables given by gauge fields given as differential 1-forms $A=A_{\mu}^AG_Adx^{\mu}$ valued on the gauge group $\mathcal{G}$ algebra with generators $G_A$, with $\mu =0,...,D$, with $D$ even, the embedding coordinates of each brane 
$X_{(d+1)}^{\mu}(\xi _{(d+1)}^i)$, where $\xi _{(d+1)}^i$, 
with $i=0,...,d$, with $d$ even, are the coordinates of the $d$ brane, and the intrinsic metrics defined at the boundaries of the branes $\gamma _{(d)rs}$, with $r,s=0,...,d-1$. It was argued in \cite{Mora:2021wmz} that only the two-dimensional metric $\gamma _{(2)rs}$ does appear, as it is a non dynamical auxiliary variable. The space-time in which 
the branes are embedded can be considered itself as a brane, and is itself described by an action of 
the same mathematical form as the actions of the branes, therefore 
its coordinates are $\xi _{(D+1)}^{\mu}\equiv x^{\mu}$. While the branes themselves may or may not have boundaries, the boundaries of the branes have no boundary. 
The actions considered, introduced in Ref.\citep{Mora-Nishino20001}, are given as a sum of terms of diverse dimensions corresponding to the various branes and the space-time in which these branes are embedded\footnote{The sum below starts in $n=1$, and not in $n=0$, because
we are considering groups for which $<F>=<\overline{F}>=0$. }
\begin{equation}
S=\sum _{n=1}^{\frac{D}{2}}S_{2n+1}^{(Trans)} +S_{2}^{(Kin)},
\end{equation}
where the Transgression action part is given by
\begin{equation}
S_{d+1}^{(Trans)}=k _{d+1}\int _{\mathcal{M}^{d+1}} \mathfrak{T}_{d+1},
\end{equation}
with the Transgression forms given by 
\begin{equation}
\mathfrak{T}_{d+1}(A,\overline{A})=[\frac{d}{2}+1]\int_0^1dt<\Delta AF_t^{\frac{d}{2}}>.
\end{equation}
There $\Delta A=A-\overline{A}$, $A_t=tA+(1-t)\overline{A}$ and $F_t=dA_t+A_t^2$. The brackets $<...>$ denote symmetrized (super)traces on the algebra of the gauge (super)group, corresponding to symmetric invariant tensors $g_{A_1...A_{d}}\equiv <G_{A_1}...G_{A_{d}}>$. 
The definition above implies $d\mathfrak{T}_{d+1}(A,\overline{A}) =F(A)^{\frac{d}{2}+1}-\overline{F}(\overline{A})^{\frac{d}{2}+1}$.\\
The Kinetic part of the action has support on the boundaries of the 
2-branes, if such boundaries exist, and is of the form
\begin{equation}
S_{2}^{(Kin)}=\pm\frac{k_2}{2}\int _{\mathcal{M}^2\equiv \partial \mathcal{M}^{3}}d^2\xi _{(2)}
\sqrt{-\gamma } ~\gamma ^{ij}<\Delta A_i\Delta A_j>,
\end{equation}
where $\Delta A_i=\Delta A_{\mu}^A\frac{\partial X_{(2)}^{\mu}}{\partial \xi _{(2)}^i}G_A$.


Some comments on the motivations of these models, as presented in \cite{Mora-Nishino20001,Mora:2021wmz} and references therein, may be helpful.
As mentioned in the Introduction, I regard gauge invariance as a fundamental aspect of physical law with deep roots on quantum mechanics.
The field strength or curvature $F$ is the basic covariant object in gauge theory, and in the absence of a space-time metric the basic
invariant objects are given by traces, or contractions with invariant tensors, of powers of $F$. These traces must be symmetrized traces, 
as they are contracted with the symmetric power of $F$, and are themselves not suited to be used as actions, as they are total 
derivatives locally. What is possible is to use Chern-Simons forms as actions 
\cite{Achucarro:1987vz,Chamseddine:1990gk, Chamseddine:1989nu, MuellerHoissen:1990vf, Banados:1993ur, Banados:1996hi, Zanelli:2012px} , 
which do satisfy $STr[F^r]=d\Omega _{2r-1}$. In a series of
papers \cite{MOTZ-2004,MOTZ-2006,Mora20141} we shown that the replacement of Chern-Simons-forms by their globally defined counterparts,
the well known to mathematicians "Transgression forms", is required to have a well defined action principle and finite conserved charges 
and thermodynamic quantities for Chern-Simons gravity. Transgression forms do depend on two sets of gauge fields, $A$
and $\overline{A}$, instead of just one set $A$, as Chern-Simons forms do. If we set $\overline{A}=0$ in the Transgression 
form we obtain the Chern-Simons form. The role of the second gauge field $\overline{A}$ was discussed 
in \cite{MOTZ-2004,MOTZ-2006,Mora20141}, and two consistent possibilities were considered: either $\overline{A}$ is 
an inert fixed background (ideally satisfying the classical equations of motion), or it is a second fully dynamical gauge field.
For the case of $\overline{A}$ being a fixed background we again identified two possibilities, roughly corresponding to the
"background substraction" and to the "counterterms" methods of regularization.\\
The way to the introduction of extended objects in \cite{Mora-Nishino20001} was inspired on the definition of 
Chern characters as a formal sum of differential forms of different orders 
$STr\{ \exp [i\frac{F}{2\pi }]\}= 1+ ch _2[F]+ ch _4[F]+ ...=1-\frac{1}{(2\pi )^2 2! }STr[F^2]+ \frac{1}{(2\pi )^4 4!}STr[F^4]+ ...$,
which suggested the corresponding sum of Chern-Simons forms of different orders as a candidate action, 
to be integrated over manifolds of the suitable dimension. We felt that these expressions for the action did embed the concept of 
giving physical content to the differential structure, and did provide another justification for the introduction of extended objects 
in fundamental physics. Reference \cite{Dixon:1991xz} suggested to us to use Transgression forms instead of Chern-Simons forms, 
as it yields a gauge invariant theory without the need to fix the gauge in the mobile boundaries of the branes.\\
Kinetic terms were added at the boundaries of the branes in ref.\cite{Mora-Nishino20001}, both of the standard and Born-Infeld form.
That was motivated by the reduction to standard String Theory models in the pure gauge case, as discussed in that reference.
For quite some time I looked for a way to obtain that kinetic term from the natural boundary term of the Transgression form, 
perhaps trough a proper choice of $\overline{A}$, without having to put it by hand, and I did not find a simple and natural way to do it. 
In ref.\cite{Mora:2021wmz} I concluded that, if the models were to be pure gauge systems, the world-volume metrics for higher dimensional
branes and the corresponding kinetic terms should vanish, while the world-volume metric for the two dimensional 
boundary of the two brane could be kept, as that metric in that case can be eliminated, and therefore it can 
be considered as an auxiliary non dynamical field.\\
It is worthwhile to note that in standard String Theory one starts with the kinetic term, namely the Nambu-Goto action, 
and the WZW term comes afterwards, through the requirement of kappa symmetry for the Superstring. In our approach the Chern-Simons or,
in the pure gauge case, the WZW term comes first, then the kinetic term is added afterwards.\\
The relative coefficient between the kinetic term and the three dimensional transgression is fixed, and equal to $\pm \frac{1}{2}$.
This choice of the relative coefficient is related to \cite{Witten-bosonization} the decoupling of the right and left movers, 
to bosonization and to the vanishing of the beta function, as it was discussed in \cite{Mora:2021wmz}. Before that, in the 
supersymmetric case, the relative coefficient was justified in ref.\cite{Mora-Nishino20001} as the one required by kappa symmetry.\\
Summarizing, in these models we have the auxiliary space $\mathcal{M}^{p+2}$ in which $<F^{\frac{p}{2}+1}>$ lives, 
then the physical world-volume of the p-brane $\mathcal{M}^{p+1}\equiv \partial \mathcal{M}^{p+2} $, 
and then the boundary 
of the brane $\mathcal{M}^{p}\equiv \partial \mathcal{M}^{p+1} $, which for 2-branes are strings \cite{Mora:2021wmz}.
These brane boundaries are assumed to be without boundary themselves.

\section{Quantization of constants for Chern-Simons branes}

In this Section I show that the quantized constants corresponding to extended objects of different 
dimensions do satisfy certain relationships, if we assume that the trace factorizes 
as $<F^{\frac{D}{2}+1}>=C~<F^r><F^s>$ with $r$ and $s$ positive integers such 
that $r+s=\frac{D}{2}+1=n+1$ and $C$ some constant.\\
The standard quantization condition in the constants of the action 
\cite{Zanelli:1994ti, Mora:2021wmz} is a consequence of the fact that the Chern-Simons forms $\Omega _{2n+1}$ on a manifold with boundary satisfy 
$\int _{\partial \mathcal{M}} \Omega _{2n+1}=\int _{\mathcal{M}}\Omega _{2n+2}$. If the boundary $\partial\mathcal{M}$ shrinks
to zero the probability amplitude for this process should be equal to one, or 
$\exp [\frac{i}{\hbar}\int _{\partial \mathcal{M}} \Omega _{2n+1}]=\exp [\frac{i}{\hbar}\int _{ \mathcal{M}} \Omega _{2n+2}]=1$ if
$\partial\mathcal{M}\rightarrow 0$, and therefore $\frac{1}{\hbar}\int _{ \mathcal{M}} \Omega _{2n+2}=2\pi N_{n+1}$ with $N_{n+1}$ integer.
If the invariant polynomials are given by $\Omega _{2n+2}=k_{n+1}<F^{n+1}>$, and 
$\int _{\mathcal{M}^{d+2}}<F^{n+1}>=\mathcal{N}_{n+1}=integer$ as a result of an index theorem \footnote{In this section and in what follows we will take $<F ^r>=constant \times STr[F^r]$ as a symmetrized invariant trace normalized to be integrated to an integer in a closed manifold, 
$\int _{\mathcal{M}^{2r}}<F^{r}>=\mathcal{N}_{r}=integer$, trough an index theorem and a proper choice of the multiplicative constant.}, 
then $\frac{k_{n+1}}{\hbar}\mathcal{N}_{n+1}=2\pi N_{n+1}$. While $N_{n+1}$ and $\mathcal{N}_{n+1}$ can vary depending on how do 
we extent de boundary, $k_{n+1}$ is chosen once and forever, which implies that $\frac{k_{n+1}}{2\pi\hbar}=z_{n+1}=integer$.
 
If the trace factorizes as $<F^{n+1}>=C<F^r><F^s>$, with the dimension of space-time equal to $2n+1=D+1$, then consistency
in the case of product manifolds implies 
$\int _{\mathcal{M}^{2n+2}}<F^{\frac{D}{2}+1}>=C\int _{\mathcal{M}^{2r}}<F^r>\int _{\mathcal{M}^{2s}}<F^s>$, or 
$\int _{\mathcal{M}^{2n+2}}\Omega _{2n+2}=C\frac{k_{n+1}}{k_rk_s}\int _{\mathcal{M}^{2r}}\Omega _{2r}\int _{\mathcal{M}^{2s}}\Omega _{2s}$.
Taking in account the quantization condition on each integral we get 
$2\pi \hbar N_{n+1}= C\frac{k_{n+1}}{k_rk_s}~2\pi\hbar N_r~2\pi\hbar N_s$, where $N_{n+1}$, $N_r$ and $N_s$ integers. Using
$\frac{k_{n+1}}{2\pi\hbar}=z_{n+1}$, $\frac{k_{r}}{2\pi\hbar}=z_{r}$ and $\frac{k_{s}}{2\pi\hbar}=z_{s}$, where $z_{n+1}$, $z_r$
and $z_s$ are integers. It follows $N_{n+1}= C\frac{z_{n+1}}{z_r z_s} N_r N_s$. 
Considering that the $N$'s depend on the manifolds considered, 
but the $k$'s, and therefore the $z$'s are given once and forever, it must be $C\frac{z_{n+1}}{z_rz_s}=integer$.

\section{Charges}

In the case of standard p-form fields, one has a potential $A_p$ and its field strength $F_{p+1}$ in a $D+1$ dimensional space-time,
and can define electric and magnetic charges of the schematic forms 
$q_E=\int _{\mathcal{M}^{D-p-2}}~^*F_{p+1}$ and $q_M=\int _{\mathcal{M}^{p+1}}F_{p+1}$ respectively.

In our framework, and in the case in which the trace factorizes as $<F(A)^{n+1}>=C~<F(A)^{\frac{p}{2}+1}>_I<F(A)^{n-\frac{p}{2}}>_{II}$,
we can define, if we are considering p-branes coupled to the gauge field, the "electric" charge 
$q_E=\int _{\mathcal{M}^{2n-p}}<F(A)^{n-\frac{p}{2}}>_{II}$ and the "magnetic" charge 
$q_M=\int _{\mathcal{M}^{p+2}}<F(A)^{\frac{p}{2}+1}>_I$. These charges are in principle quantized, that is $q=\mathcal{N}=integer$.\\
The field configuration satisfy 
$q_E=\int _{\mathcal{M}^{2n-p}\equiv\partial \mathcal{M}^{2n-p+1}}<F(A)^{n-\frac{p}{2}}>_{II}=\int _{\mathcal{M}^{2n-p+1}}d<F(A)^{n-\frac{p}{2}}>_{II}$. 
The last relationship implies $d<F(A)^{n-\frac{p}{2}}>_{II}=q_E\delta _{\mathcal{M}^{p}}$, where $\mathcal{M}_p\equiv\partial \mathcal{M}^{p+1}$ and
$\delta _{\mathcal{M}^{p}}$ is a Dirac delta in the coordinates normal to $\mathcal{M}^p$. Notice that the charge is necessarily singular, as in the absence of any singularity $d<F^r>=<D(F^r)>=r<F^{r-1}DF>=0$, in virtue of the Bianchi identity.

\section{Enhanced gauge invariance for Chern-Simons branes}

As shown in Ref.\cite{Mora-Nishino20001}, the action of Section 3 is gauge invariant if both gauge fields $A$ and $\overline{A}$ have the same gauge transformation,
as $F$, $\overline{F}$ and $\Delta A$, and hence $F_t$, are gauge covariant and $\gamma _{ij}$ is invariant in that case. However, under
a gauge transformation of only one of the gauge fields ($\delta A\neq 0$ and $\delta\overline{A}=0$), the actions change by terms at the boundaries of the branes or the space-time. 
This is what happens for instance in the pure Chern-Simons case, where $\overline{A}$ is just zero everywhere.

In this Section I show that it is possible to have a gauge invariant action 
even in the case that only one of the gauge fields change, if the trace factorizes,
by using the inflow mechanism.

The Transgression satisfies 
$<F(A)^{n+1}>-<\overline{F}(\overline{A})^{n+1}>=d\mathfrak{T}_{2n+1}(A,\overline{A})$.
If the trace factorizes as $<F(A)^{n+1}>=C<F(A)^{\frac{p}{2}+1}>_I<F(A)^{n-\frac{p}{2}}>_{II}$,
where $p$ is an even integer and $<...>_I$ and $<...>_{II}$ are the traces into which the original trace factorizes,
then it can be shown that\footnote{We could as well have written a similar expression with the roles of the traces $I$ and $II$
interchanged.} 
\begin{eqnarray}
<F(A)^{n+1}>-<\overline{F}(\overline{A})^{n+1}>= \nonumber\\
=C\{ d\mathfrak{T}^{(I)}_{p+1}(A,\overline{A})
[<F^{n-\frac{p}{2}}(A)>_{II}+<\overline{F}^{n-\frac{p}{2}}(\overline{A})>_{II}]+\nonumber\\
+<\overline{F}^{\frac{p}{2}+1}(\overline{A})>_I<F^{n-\frac{p}{2}}(A)>_{II}-
<F^{\frac{p}{2}+1}(A)>_I<\overline{F}^{n-\frac{p}{2}}(\overline{A})>_{II}\}
\end{eqnarray}
This expression simplifies notably if we assume $\overline{F}=0$, that is if $\overline{A}$ is a background 
or a "vacuum"\footnote{In pure Chern-Simons theories $\overline{A}=0$ and $\overline{F}=0$, but in that case other problems appear,
as for instance not having a well defined action principle \cite{MOTZ-2004,MOTZ-2006}. A more suitable background for theories with
Anti de Sitter gauge group is anti de Sitter space-time.}
, in which case it becomes
\begin{equation}
<F(A)^{n+1}>=C d\mathfrak{T}^{(I)}_{p+1}(A,\overline{A})
<F^{n-\frac{p}{2}}(A)>_{II}.
\end{equation}
This imply $<F(A)^{n+1}>=d[C \mathfrak{T}^{(I)}_{p+1}(A,\overline{A})
<F^{n-\frac{p}{2}}(A)>_{II}]$, because $d<F(A)^{n-\frac{p}{2}}>_{II}=0$ in the absence of a p-brane, which in turn means
\begin{equation}
\int _{\mathcal{M}^{2n+2}} <F(A)^{n+1}>=\int _{\mathcal{M}^{2n+1}\equiv \partial \mathcal{M}^{2n+2}}[C \mathfrak{T}^{(I)}_{p+1}(A,\overline{A})
<F^{n-\frac{p}{2}}(A)>_{II}].
\end{equation}
While the previous expressions do not hold in the presence of a fundamental p-brane, we will consider them as the definition 
of the Transgression if the trace factorizes and $\overline{F}=0$, that is 
$\mathfrak{T}_{2n+1}(A,\overline{A})\equiv C\mathfrak{T}^{(I)}_{p+1}(A,\overline{A})
<F^{n-\frac{p}{2}}(A)>_{II} $.

The gauge variation of the transgression is 
\begin{equation*}
\delta \mathfrak{T}_{2n+1}(A,\overline{A})=C\delta [\mathfrak{T}^{(I)}_{p+1}(A,\overline{A})]
<F^{n-\frac{p}{2}}(A)>_{II}=Cd[\Omega^{1(I)}_p(A,\overline{A})] <F^{n-\frac{p}{2}}(A)>_{II}, 
\end{equation*}
after taking in account 
that $\delta <F^{n-\frac{p}{2}}(A)>_{II}=0$, and introducing 
the gauged Wess-Zumino-Witten functional $\Omega^{1(I)}_p(A,\overline{A})$
associated to the trace $I$. Integration by parts yields 
\begin{equation*}
\delta \mathfrak{T}_{2n+1}(A,\overline{A})=-C\Omega^{1(I)}_p(A,\overline{A})d[ <F^{n-\frac{p}{2}}(A)>_{II}]+d[C\Omega^{1(I)}_p(A,\overline{A}) <F^{n-\frac{p}{2}}(A)>_{II}]. 
\end{equation*}
In the first term of the second member we need that in presence of a fundamental, singular, p-brane 
$d[ <F^{n-\frac{p}{2}}(A)>_{II}]=q_E\delta _{\mathcal{M}^{p}}$
, where $\delta _{\mathcal{M}^{p}}$ is a Dirac delta on the coordinates 
normal to the boundary of the p-brane, as in subsection 2.3,
and $q_E$ the charge of the brane, located at its boundary. 
Then
\begin{eqnarray}
\int _{\mathcal{M}^{2n+1}}\delta\mathfrak{T}_{2n+1}=\nonumber\\
=-C\int _{\mathcal{M}^{2n+1}}\Omega^{1(I)}_p(A,\overline{A})d[ <F^{n-\frac{p}{2}}(A)>_{II}]
+\int _{\mathcal{M}^{2n+1}}d[C\Omega^{1(I)}_p(A,\overline{A}) <F^{n-\frac{p}{2}}(A)>_{II}]=\nonumber\\
=-q_E C\int _{\mathcal{M}^{p}}\Omega^{1(I)}_p(A,\overline{A})+space-time~boundary~term. \nonumber
\end{eqnarray}
We will assume the space-time boundary term vanish, either because of the fall-off behaviour of the fields 
or because the space-time has no boundary.\\
On the other hand, the gauge variation of the p-brane action is 
\begin{equation*}
k_{p+1}\int _{\mathcal{M}^{p+1}}\delta\mathfrak{T}_{p+1}=
k_{p+1}\int _{\mathcal{M}^{p}\equiv\partial\mathcal{M}^{p+1}}\Omega^{1(I)}_p(A,\overline{A}). 
\end{equation*}
Comparing both expressions we see that these variations cancel if $k_{p+1}=q_E C k_{2n+1}$. This cancellation only works for 
one of the possible values of $q_E$, or equivalently the relationship between the coefficients must be corrected for different charges.

What remains is to consider is the variation of the two dimensional kinetic term. We have
$\delta S_{2}^{(Kin)}=\pm\frac{k_2}{2}\int _{\mathcal{M}^2\equiv \partial \mathcal{M}^{3}}d^2\xi _{(2)}
\delta [\sqrt{-\gamma}\gamma ^{ij}<\Delta A_i\Delta A_j>]$. For this expression to be invariant it must be
$0=\delta [\sqrt{-\gamma}\gamma ^{ij}<\Delta A_i\Delta A_j>]=
\delta [\sqrt{-\gamma}\gamma ^{ij}]<\Delta A_i\Delta A_j>+\sqrt{-\gamma}\gamma ^{ij}\delta[<\Delta A_i\Delta A_j>]$.
The gauge variation of $\Delta A$, if only $A$ varies, is $\delta\Delta A=\delta A=D\lambda$, where $\lambda $ is the
infinitesimal gauge parameter. For the required vanishing of the kinetic term variation we must assign the auxiliary metric in the world-sheet
a transformation such that the above condition is valid. To see that it is possible, we may for instance use the Weyl invariance of this term 
to fix the metric determinant to minus one, in which case the condition is just $\delta[\gamma ^{ij}<\Delta A_i\Delta A_j>]=0$.
That condition is consistent, even tough gauge transformations are off-shell, with the equations of motion, that tell us \cite{Mora:2021wmz} that the auxiliary metric is $\gamma _{ij}=<\Delta A_i\Delta A_j>$.\\
A dual version of the previous considerations is possible, with the roles of the traces $<...>_I$ and $<...>_{II}$ reversed.
In that case instead of a p-brane we have a $(2n-p-1)$-brane.

\section{Concrete examples of trace factorization}

Following Ref.\cite{Green:1984sg}, in the case of the group $SO(n)$ one has the trace "tr" in the fundamental representation 
of dimension $n$ and the trace "Tr" in the adjoint representation. If we consider the indices $a,b,c,d=1,...,n$,
then the field strength in the adjoint representation is 
$F_{ab,cd}=\frac{1}{2}[F_{ac}\delta _{bd}-F _{bc}\delta _{ad}-F_{ad}\delta _{bc}+F_{bd}\delta _{ac}]$, where the antisymmetric 
$n\times n$ matrix $F_{ab}$ is $F$ in the fundamental representation. Taking traces of powers of $F$ using this 
expression it results, for the first even powers of $F$ \footnote{The traces of any odd power of $F$ vanish, that is 
$Tr F^{2n+1}=tr F^{2n+1}=0$ for $n=0,1,...$. },
\begin{eqnarray}
Tr F^2=(n-2)~tr F^2\\
Tr F^4=(n-8)~tr F^4+3~(tr F^2)^2\\
Tr F^6=(n-32)~tr F^6+15 ~tr F^2tr F^4.
\end{eqnarray}        
We see from these expressions that $Tr F^4$ factorizes if $n=8$ and that $Tr F^6$ factorizes if $n=32$.
We can also solve to write the factorization condition in terms of only adjoint traces as
\begin{eqnarray}
Tr F^4=\frac{1}{12}(Tr F^2)^2~~~if~~~n=8 \\
Tr F^6=Tr F^2\left[\frac{1}{48}Tr F^4 -\frac{1}{1440}(Tr F^2)^2\right]~~~if~~~n=32
\end{eqnarray}

For a de Sitter $SO(d,1)$ or Anti de Sitter $SO(d-1,2)$ space-time groups the relationship given above holds, 
with the Kronecker delta $\delta _{ab}$ replaced by the Minkowski 
invariant tensor $\eta _{ab}$ with the suitable signature, that is
$F_{ab,cd}=\frac{1}{2}[F_{ac}\eta _{bd}-F _{bc}\eta _{ad}-F_{ad}\eta _{bc}+F_{bd}\eta _{ac}]$.
Raising the indices $a,b$ we get 
$F ^{ab}_{~~cd}=\frac{1}{2}[F^{a}_{~c}\delta ^{b}_{~d}-F ^{b}_{~c}\delta ^{a}_{~d}-F^{a}_{~d}\delta ^{b}_{~c}+F^{b}_{~d}\delta ^{a}_{~c}]$.
Taking traces of powers of $F$ in both sides it follows that the relationships between the traces hold for these groups, and that factorization occurs in the
same cases.\\
The cases $n=8$ and $n=32$ are again cases of interest.
The case of the AdS group $SO(6,2)$ corresponds to a seven dimensional (6+1) theory with dual 2-branes, 
or in the pure gauge case a 5-brane background and dual strings. In this case the number 
of generators and therefore gauge fields is exactly the required for the vielbein (7) and spin connection ($\frac{7\times 6}{2}=21$) of a seven dimensional theory,
that is $\frac{8\times 7}{2} =7+\frac{7\times 6}{2}=28$.\\
The case of the AdS group $SO(30,2)$ corresponds to an eleven dimensional (10+1) theory with 2-branes and dual 6-branes. 
In this case the group has $\frac{32\times 31}{2}=496$ generators, and therefore the same number of gauge potentials, of which 11 are identified as the vielbein and  $\frac{11\times 10}{2}=55$ as the spin connection. The remaining $430=496-55-11$ gauge fields correspond to 
internal local symmetries. The gauge fields associated to these local internal symmetries and the vielbein and spin connection do mix under general gauge transformations.\\
Writing the actions for these theories is immediate, but to write the explicit solutions for the extended objects, the corresponding gauge fields, and the background in the static case is an interesting problem that will be addressed in future work, as are other aspects of these models.

\section{Discussion and Conclusions}

The original motivation of this work are Refs.\cite{Duff:1995wd,Witten:1995em}, both foundational articles on M-theory
\cite{Hull:1994ys,Witten:1995ex}, and the 
goal of the present program of research is indeed a non perturbative definition of that theory, as already stated in 
Ref.\cite{Mora-Nishino20001}. 
My strategy is to take the wealth of previous results as 
suggestions of the existence and properties of that theory, but to reinterpret those results in the framework put forward in Refs.\cite{Mora-Nishino20001,Mora:2021wmz} and this article.
To that end several further steps must be taken.\\
We need to consider the fermionic case, introduced in a Supersymmetric way, 
via a suitable choice of the supergroup and invariant supertrace. A good candidate supergroup could be $OSp(32,1)$ 
or some of its contractions,
as it has been suggested in Refs.\cite{Troncoso1997va, Troncoso1998ng,Horava:1997dd,Izaurieta:2011fr,Mora-Nishino20001}, 
while the invariant tensor could be the invariant supertrace. 
The theory we are looking for should be eleven dimensional, with 2-branes and 5-branes. In Ref.\cite{Mora-Nishino20001} we suggested a possible candidate, given by eleven dimensional Chern-Simons supergravity with gauge group $OSp(32,1)$ and Chern-Simons 2-branes and 6-branes. 
Our 2-branes are not the standard ones, as they have the kinetic term located at the boundary, and reduce to 1-branes or strings 
in the pure gauge case.Those fundamental branes may give a long wavelength theory similar to the standard 2-brane. 
In an eleven dimensional theory 
we should have 6-branes, and they have no kinetic term at all\footnote{The lack of a kinetic term implies that no 
perturbative expansion is possible. That may be the reason for the often mentioned fact that 5-branes are purely quantum and non perturbative
objects.}, but the 5-branes of M-theory may be the ones to which they reduce in the pure gauge case. 
The background theory is Chern-Simons supergravity in eleven dimensions, which has been shown \cite{Izaurieta:2011fr} to yield a theory remarkable similar to standard eleven dimensional supergravity \cite{Cremmer:1978km} in certain limit, plus corrections. It remains to be found, if possible, a suitable invariant tensor that would exactly yield standard eleven dimensional 
supergravity in a limit, plus corrections. A guide in that search may be the term $A_3F_4F_4$ in the action of that theory, mentioned in Section 2.3, as in our framework $A_3$ should be a Chern-Simons form and $F_4=C<F^2>$ schematically, and the form of that term suggest 
a double factorization of the trace for at least part of the algebra.\\
The factorization of the trace and the ensuing enhanced gauge invariance should be a criterion 
to select the right dimension, group and trace. More symmetry means that the true degrees of freedom 
of the theory are less than the apparent ones, and a description that is in essence simpler than it seems\footnote{That may be a general property of fundamental physical law, in the spirit of Wheeler's thoughts cited above, perhaps similar to the concept of $Y\overline{u}gen$ 
in Japanese aesthetics.}.\\
Anomaly cancellation arguments should also play a role in selecting the right theory, as they did in String Theory \cite{Green:1984sg}.
At the quantum level the models discussed in this work should in principle be anomalous, as for instance there would in general be a world sheet Weyl anomaly for the boundaries of 2-branes/strings, as in standard String Theory, presumably vanishing in 26 dimensions.
Also the chiral bosons introduced by the choice of the relative coefficient of the two parts of the action of the 2-brane should generate 
a gravitational anomaly owing to their coupling to the metric $\gamma _{ij}$\footnote{A chiral scalar 
could be thought 
as a $p=0$ chiral p-form field, which would give \cite{Alvarez-Gaume:1983ihn} an anomaly polynomial 
of $I_A=\frac{1}{48}tr [R_{(2)}^2]$ for each chiral boson, where $R_{(2)}$
is the curvature associated to the metric $\gamma _{ij}$.}. In theories with fermions one would expect anomalies associated with the
even dimensional boundaries of the branes, and a much more constrained and intricate pattern of anomaly cancellations, that would severely restrict viable theories.\\
I wonder if, in case an anomaly free model is found, the action of that model would not be itself the quantum effective 
action \cite{Jona-Lasinio:1964zvf}, receiving no quantum corrections. This conjecture is motivated by the Adler-Bardeen theorem 
\cite{Adler:1969er}\footnote{See also Ref.\cite{Zee:2003mt}, Ch.IV.7 for a clear and pedagogical exposition.}, and the fact that the mathematical structure of our action is just the same than the one of Anomalies. 
This may seem surprising, as it is commonly assumed that for interacting theories the quantum effective action must be non local, but I am not aware of any general result or theorem in that sense\footnote{I acknowledge helpful correspondence with Prof. K. Scharnhorst on this matter. Also I found his work, Refs.\cite{Scharnhorst:1993bk,Scharnhorst:2023spc} relevant on this topic.}, and also what seems a non local theory in some dimension may come from a local theory in a higher dimension, as shown by Kaluza-Klein theories, and wormholes.\\
Another interesting question has to do with topological quantum phases in these models, in the line of 
Refs.\cite{Wilczek:1983cy,Wu:1984kd,Nepomechie1984,Wu:1988py},
and the implications of this phases, among other things, for the statistics.\\
Finally, Chern-Simons models have no intrinsic scale, but one is introduced in the case of AdS groups and they 
supersymmetric generalizations
\cite{Achucarro:1987vz,Chamseddine:1990gk, Chamseddine:1989nu, MuellerHoissen:1990vf, Banados:1993ur, Banados:1996hi, Zanelli:2012px} by 
identifying part of the gauge potentials as the vielbein $A^a_{\mu}=\frac{1}{l}e^a_{\mu}$, where $l$ is such constant. I think there is more to be said and done about that question\footnote{In the same vein of Schrodinger's cat problem in Quantum Mechanics, we may call
this Galileo's dog problem, after the classic discussion and drawing on scale invariance not being a symmetry of Nature in Galileo's book {\it Discourses and mathematical demonstrations relating to two new Sciences}.}.\\

\acknowledgments

While working in the present article I received financial support from the {\it Sistema Nacional de Investigadores} (SNI), of the
{\it Agencia Nacional de Investigaci\'on e Innovaci\'on} (ANII) of Uruguay.

\bibliography{mora-V-2023-bibfyle}

\begin{thebibliography}{10}
\expandafter\ifx\csname url\endcsname\relax
  \def\url#1{\texttt{#1}}\fi
\expandafter\ifx\csname urlprefix\endcsname\relax\def\urlprefix{URL }\fi
\expandafter\ifx\csname href\endcsname\relax
  \def\href#1#2{#2} \def\path#1{#1}\fi

\bibitem{Mora-Nishino20001}
P.~Mora, H.~Nishino, {Fundamental extended objects for Chern–Simons
  supergravity}, Physics Letters B 482 (2000) 222--232.
\newblock \href {http://arxiv.org/abs/hep-th/0002077}
  {\path{arXiv:hep-th/0002077}}, \href
  {https://doi.org/10.1016/S0370-2693(00)00535-9}
  {\path{doi:10.1016/S0370-2693(00)00535-9}}.

\bibitem{Achucarro:1987vz}
A.~Achucarro, P.~Townsend, {A Chern-Simons Action for Three-Dimensional anti-De
  Sitter Supergravity Theories}, Phys. Lett. B 180 (1986) 89.
\newblock \href {https://doi.org/10.1016/0370-2693(86)90140-1}
  {\path{doi:10.1016/0370-2693(86)90140-1}}.

\bibitem{Chamseddine:1990gk}
A.~H. Chamseddine, {Topological gravity and supergravity in various
  dimensions}, Nucl. Phys. B 346 (1990) 213--234.
\newblock \href {https://doi.org/10.1016/0550-3213(90)90245-9}
  {\path{doi:10.1016/0550-3213(90)90245-9}}.

\bibitem{Chamseddine:1989nu}
A.~H. Chamseddine, {Topological Gauge Theory of Gravity in Five-dimensions and
  All Odd Dimensions}, Phys. Lett. B 233 (1989) 291--294.
\newblock \href {https://doi.org/10.1016/0370-2693(89)91312-9}
  {\path{doi:10.1016/0370-2693(89)91312-9}}.

\bibitem{MuellerHoissen:1990vf}
F.~Mueller-Hoissen, {From Chern-Simons to Gauss-Bonnet}, Nucl. Phys. B 346
  (1990) 235--252.
\newblock \href {https://doi.org/10.1016/0550-3213(90)90246-A}
  {\path{doi:10.1016/0550-3213(90)90246-A}}.

\bibitem{Banados:1993ur}
M.~Banados, C.~Teitelboim, J.~Zanelli, {Dimensionally continued black holes},
  Phys. Rev. D 49 (1994) 975--986.
\newblock \href {http://arxiv.org/abs/gr-qc/9307033}
  {\path{arXiv:gr-qc/9307033}}, \href {https://doi.org/10.1103/PhysRevD.49.975}
  {\path{doi:10.1103/PhysRevD.49.975}}.

\bibitem{Banados:1996hi}
M.~Banados, R.~Troncoso, J.~Zanelli, {Higher dimensional Chern-Simons
  supergravity}, Phys. Rev. D 54 (1996) 2605--2611.
\newblock \href {http://arxiv.org/abs/gr-qc/9601003}
  {\path{arXiv:gr-qc/9601003}}, \href
  {https://doi.org/10.1103/PhysRevD.54.2605}
  {\path{doi:10.1103/PhysRevD.54.2605}}.

\bibitem{Zanelli:2012px}
J.~Zanelli, {Chern-Simons Forms in Gravitation Theories}, Class. Quant. Grav.
  29 (2012) 133001.
\newblock \href {http://arxiv.org/abs/1208.3353} {\path{arXiv:1208.3353}},
  \href {https://doi.org/10.1088/0264-9381/29/13/133001}
  {\path{doi:10.1088/0264-9381/29/13/133001}}.

\bibitem{Mora:2021wmz}
P.~Mora, {Actions, equations of motion and boundary conditions for Chern-Simons
  branes}, Phys. Lett. B 819 (2021) 136428.
\newblock \href {http://arxiv.org/abs/2105.08863} {\path{arXiv:2105.08863}},
  \href {https://doi.org/10.1016/j.physletb.2021.136428}
  {\path{doi:10.1016/j.physletb.2021.136428}}.

\bibitem{Dixon:1991xz}
J.~A. Dixon, M.~Duff, E.~Sezgin, {The Coupling of Yang-Mills to extended
  objects}, Phys. Lett. B 279 (1992) 265--271.
\newblock \href {http://arxiv.org/abs/hep-th/9201019}
  {\path{arXiv:hep-th/9201019}}, \href
  {https://doi.org/10.1016/0370-2693(92)90391-G}
  {\path{doi:10.1016/0370-2693(92)90391-G}}.

\bibitem{Green:1989nn}
M.~B. Green, {Supertranslations, Superstrings and \{Chern-Simons\} Forms},
  Phys. Lett. B 223 (1989) 157--164.
\newblock \href {https://doi.org/10.1016/0370-2693(89)90233-5}
  {\path{doi:10.1016/0370-2693(89)90233-5}}.

\bibitem{Troncoso1997va}
R.~Troncoso, J.~Zanelli, {New gauge supergravity in seven-dimensions and
  eleven-dimensions}, Phys. Rev. D 58 (1998) 101703.
\newblock \href {http://arxiv.org/abs/hep-th/9710180}
  {\path{arXiv:hep-th/9710180}}, \href
  {https://doi.org/10.1103/PhysRevD.58.101703}
  {\path{doi:10.1103/PhysRevD.58.101703}}.

\bibitem{Troncoso1998ng}
R.~Troncoso, J.~Zanelli, {Gauge supergravities for all odd dimensions}, Int. J.
  Theor. Phys. 38 (1999) 1181--1206.
\newblock \href {http://arxiv.org/abs/hep-th/9807029}
  {\path{arXiv:hep-th/9807029}}, \href
  {https://doi.org/10.1023/A:1026614631617}
  {\path{doi:10.1023/A:1026614631617}}.

\bibitem{Horava:1997dd}
P.~Horava, {M theory as a holographic field theory}, Phys. Rev. D 59 (1999)
  046004.
\newblock \href {http://arxiv.org/abs/hep-th/9712130}
  {\path{arXiv:hep-th/9712130}}, \href
  {https://doi.org/10.1103/PhysRevD.59.046004}
  {\path{doi:10.1103/PhysRevD.59.046004}}.

\bibitem{Edelstein:2008ry}
J.~D. Edelstein, J.~Zanelli, {Sources for Chern-Simons theories}, in: {workshop
  on Quantum Mechanics of Fundamental Systems: the Quest for Beauty and
  Simplicity: Dedicated to Claudio Bunster on the occasion of his 60th
  birthday}, 2009, pp. 107--124.
\newblock \href {http://arxiv.org/abs/0807.4217} {\path{arXiv:0807.4217}},
  \href {https://doi.org/10.1007/978-0-387-87499-9_8}
  {\path{doi:10.1007/978-0-387-87499-9_8}}.

\bibitem{Miskovic:2009dd}
O.~Miskovic, J.~Zanelli, {Couplings between Chern-Simons gravities and
  2p-branes}, Phys. Rev. D 80 (2009) 044003.
\newblock \href {http://arxiv.org/abs/0901.0737} {\path{arXiv:0901.0737}},
  \href {https://doi.org/10.1103/PhysRevD.80.044003}
  {\path{doi:10.1103/PhysRevD.80.044003}}.

\bibitem{Edelstein:2010sx}
J.~D. Edelstein, A.~Garbarz, O.~Miskovic, J.~Zanelli, {Stable p-branes in
  Chern-Simons AdS supergravities}, Phys. Rev. D 82 (2010) 044053.
\newblock \href {http://arxiv.org/abs/1006.3753} {\path{arXiv:1006.3753}},
  \href {https://doi.org/10.1103/PhysRevD.82.044053}
  {\path{doi:10.1103/PhysRevD.82.044053}}.

\bibitem{Edelstein:2010sh}
J.~D. Edelstein, A.~Garbarz, O.~Miskovic, J.~Zanelli, {Naked Singularities,
  Topological Defects and Brane Couplings}, Int. J. Mod. Phys. D 20 (2011)
  839--849.
\newblock \href {http://arxiv.org/abs/1009.4418} {\path{arXiv:1009.4418}},
  \href {https://doi.org/10.1142/S0218271811019177}
  {\path{doi:10.1142/S0218271811019177}}.

\bibitem{Edelstein:2011vu}
J.~D. Edelstein, A.~Garbarz, O.~Miskovic, J.~Zanelli, {Geometry and stability
  of spinning branes in AdS gravity}, Phys. Rev. D 84 (2011) 104046.
\newblock \href {http://arxiv.org/abs/1108.3523} {\path{arXiv:1108.3523}},
  \href {https://doi.org/10.1103/PhysRevD.84.104046}
  {\path{doi:10.1103/PhysRevD.84.104046}}.

\bibitem{Kastikainen:2020auf}
J.~Kastikainen, {Conical defects and holography in topological AdS gravity},
  Class. Quant. Grav. 37~(19) (2020) 195010.
\newblock \href {http://arxiv.org/abs/2006.02803} {\path{arXiv:2006.02803}},
  \href {https://doi.org/10.1088/1361-6382/abac44}
  {\path{doi:10.1088/1361-6382/abac44}}.

\bibitem{Frey:2019fqz}
A.~R. Frey, {Dirac branes for Dirichlet branes: Supergravity actions}, Phys.
  Rev. D 102~(4) (2020) 046017.
\newblock \href {http://arxiv.org/abs/1907.12755} {\path{arXiv:1907.12755}},
  \href {https://doi.org/10.1103/PhysRevD.102.046017}
  {\path{doi:10.1103/PhysRevD.102.046017}}.

\bibitem{Ertem:2012qv}
U.~Ertem, O.~Acik, {Couplings of gravitational currents with Chern-Simons
  gravities}, Phys. Rev. D 87~(4) (2013) 044052.
\newblock \href {http://arxiv.org/abs/1211.3289} {\path{arXiv:1211.3289}},
  \href {https://doi.org/10.1103/PhysRevD.87.044052}
  {\path{doi:10.1103/PhysRevD.87.044052}}.

\bibitem{Wheeler}
J.~A. Wheeler, {Hermann Weyl and the Unity of Knowledge}, American Scientist 74
  (1986) 366--375.

\bibitem{Dirac}
P.~A.~M. Dirac, The Principles of Quantum Mechanics, Oxford University Press,
  1930.

\bibitem{Weinberg}
S.~Weinberg, Dreams of a final theory, Pantheon Books, 1992.

\bibitem{Yang}
C.-N. Yang, Square root of minus one, complex phases and erwin schroedinger,
  in: C.~W. Kilmister (Ed.), Schroedinger: Centenary Celebration of a Polymath,
  Cambridge University Press, 1987, pp. 55--64.

\bibitem{Gross:1984dd}
D.~J. Gross, J.~A. Harvey, E.~J. Martinec, R.~Rohm, {The Heterotic String},
  Phys. Rev. Lett. 54 (1985) 502--505.
\newblock \href {https://doi.org/10.1103/PhysRevLett.54.502}
  {\path{doi:10.1103/PhysRevLett.54.502}}.

\bibitem{Gross:1985fr}
D.~J. Gross, J.~A. Harvey, E.~J. Martinec, R.~Rohm, {Heterotic String Theory.
  1. The Free Heterotic String}, Nucl. Phys. B 256 (1985) 253.
\newblock \href {https://doi.org/10.1016/0550-3213(85)90394-3}
  {\path{doi:10.1016/0550-3213(85)90394-3}}.

\bibitem{Gross:1985rr}
D.~J. Gross, J.~A. Harvey, E.~J. Martinec, R.~Rohm, {Heterotic String Theory.
  2. The Interacting Heterotic String}, Nucl. Phys. B 267 (1986) 75--124.
\newblock \href {https://doi.org/10.1016/0550-3213(86)90146-X}
  {\path{doi:10.1016/0550-3213(86)90146-X}}.

\bibitem{Zumino:1983rz}
B.~Zumino, Y.-S. Wu, A.~Zee, {Chiral Anomalies, Higher Dimensions, and
  Differential Geometry}, Nucl. Phys. B 239 (1984) 477--507.
\newblock \href {https://doi.org/10.1016/0550-3213(84)90259-1}
  {\path{doi:10.1016/0550-3213(84)90259-1}}.

\bibitem{Alvarez-Gaume:1983ihn}
L.~Alvarez-Gaume, E.~Witten, {Gravitational Anomalies}, Nucl. Phys. B 234
  (1984) 269.
\newblock \href {https://doi.org/10.1016/0550-3213(84)90066-X}
  {\path{doi:10.1016/0550-3213(84)90066-X}}.

\bibitem{Alvarez-Gaume:1984zlq}
L.~Alvarez-Gaume, P.~H. Ginsparg, {The Structure of Gauge and Gravitational
  Anomalies}, Annals Phys. 161 (1985) 423, [Erratum: Annals Phys. 171, 233
  (1986)].
\newblock \href {https://doi.org/10.1016/0003-4916(85)90087-9}
  {\path{doi:10.1016/0003-4916(85)90087-9}}.

\bibitem{Green:1984sg}
M.~B. Green, J.~H. Schwarz, {Anomaly Cancellation in Supersymmetric D=10 Gauge
  Theory and Superstring Theory}, Phys. Lett. B 149 (1984) 117--122.
\newblock \href {https://doi.org/10.1016/0370-2693(84)91565-X}
  {\path{doi:10.1016/0370-2693(84)91565-X}}.

\bibitem{Callan:1984sa}
C.~G. Callan, Jr., J.~A. Harvey, {Anomalies and Fermion Zero Modes on Strings
  and Domain Walls}, Nucl. Phys. B 250 (1985) 427--436.
\newblock \href {https://doi.org/10.1016/0550-3213(85)90489-4}
  {\path{doi:10.1016/0550-3213(85)90489-4}}.

\bibitem{Duff:1995wd}
M.~J. Duff, J.~T. Liu, R.~Minasian, {Eleven-dimensional origin of string-string
  duality: A One loop test}, Nucl. Phys. B 452 (1995) 261--282.
\newblock \href {http://arxiv.org/abs/hep-th/9506126}
  {\path{arXiv:hep-th/9506126}}, \href
  {https://doi.org/10.1016/0550-3213(95)00368-3}
  {\path{doi:10.1016/0550-3213(95)00368-3}}.

\bibitem{Witten:1995em}
E.~Witten, {Five-branes and M theory on an orbifold}, Nucl. Phys. B 463 (1996)
  383--397.
\newblock \href {http://arxiv.org/abs/hep-th/9512219}
  {\path{arXiv:hep-th/9512219}}, \href
  {https://doi.org/10.1016/0550-3213(96)00032-6}
  {\path{doi:10.1016/0550-3213(96)00032-6}}.

\bibitem{Dixon:1992if}
J.~A. Dixon, M.~J. Duff, J.~C. Plefka, {Putting string / five-brane duality to
  the test}, Phys. Rev. Lett. 69 (1992) 3009--3012.
\newblock \href {http://arxiv.org/abs/hep-th/9208055}
  {\path{arXiv:hep-th/9208055}}, \href
  {https://doi.org/10.1103/PhysRevLett.69.3009}
  {\path{doi:10.1103/PhysRevLett.69.3009}}.

\bibitem{Duff:1994vv}
M.~J. Duff, R.~Minasian, {Putting string / string duality to the test}, Nucl.
  Phys. B 436 (1995) 507--528.
\newblock \href {http://arxiv.org/abs/hep-th/9406198}
  {\path{arXiv:hep-th/9406198}}, \href
  {https://doi.org/10.1016/0550-3213(94)00538-P}
  {\path{doi:10.1016/0550-3213(94)00538-P}}.

\bibitem{Vafa:1995fj}
C.~Vafa, E.~Witten, {A One loop test of string duality}, Nucl. Phys. B 447
  (1995) 261--270.
\newblock \href {http://arxiv.org/abs/hep-th/9505053}
  {\path{arXiv:hep-th/9505053}}, \href
  {https://doi.org/10.1016/0550-3213(95)00280-6}
  {\path{doi:10.1016/0550-3213(95)00280-6}}.

\bibitem{Witten:1996md}
E.~Witten, {On flux quantization in M theory and the effective action}, J.
  Geom. Phys. 22 (1997) 1--13.
\newblock \href {http://arxiv.org/abs/hep-th/9609122}
  {\path{arXiv:hep-th/9609122}}, \href
  {https://doi.org/10.1016/S0393-0440(96)00042-3}
  {\path{doi:10.1016/S0393-0440(96)00042-3}}.

\bibitem{Cremmer:1978km}
E.~Cremmer, B.~Julia, J.~Scherk, {Supergravity Theory in Eleven-Dimensions},
  Phys. Lett. B 76 (1978) 409--412.
\newblock \href {https://doi.org/10.1016/0370-2693(78)90894-8}
  {\path{doi:10.1016/0370-2693(78)90894-8}}.

\bibitem{MOTZ-2004}
P.~Mora, R.~Olea, R.~Troncoso, J.~Zanelli, {Finite action principle for
  Chern-Simons AdS gravity}, Journal of High Energy Physics 06 (2004) 036.
\newblock \href {http://arxiv.org/abs/hep-th/0405267}
  {\path{arXiv:hep-th/0405267}}, \href
  {https://doi.org/10.1088/1126-6708/2004/06/036}
  {\path{doi:10.1088/1126-6708/2004/06/036}}.

\bibitem{MOTZ-2006}
P.~Mora, R.~Olea, R.~Troncoso, J.~Zanelli, {Transgression forms and extensions
  of Chern-Simons gauge theories}, Journal of High Energy Physics 02 (2006)
  067.
\newblock \href {http://arxiv.org/abs/hep-th/0601081}
  {\path{arXiv:hep-th/0601081}}, \href
  {https://doi.org/10.1088/1126-6708/2006/02/067}
  {\path{doi:10.1088/1126-6708/2006/02/067}}.

\bibitem{Mora20141}
P.~Mora, {Action Principles for Transgression and Chern-Simons AdS Gravities},
  Journal of High Energy Physics 11 (2014) 128.
\newblock \href {http://arxiv.org/abs/1407.6032} {\path{arXiv:1407.6032}},
  \href {https://doi.org/10.1007/JHEP11(2014)128}
  {\path{doi:10.1007/JHEP11(2014)128}}.

\bibitem{Witten-bosonization}
E.~Witten, {Nonabelian Bosonization in Two-Dimensions}, Commun. Math. Phys. 92
  (1984) 455--472.
\newblock \href {https://doi.org/10.1007/BF01215276}
  {\path{doi:10.1007/BF01215276}}.

\bibitem{Zanelli:1994ti}
J.~Zanelli, {Quantization of the gravitational constant in odd dimensional
  gravity}, Phys. Rev. D 51 (1995) 490--492.
\newblock \href {http://arxiv.org/abs/hep-th/9406202}
  {\path{arXiv:hep-th/9406202}}, \href
  {https://doi.org/10.1103/PhysRevD.51.490}
  {\path{doi:10.1103/PhysRevD.51.490}}.

\bibitem{Hull:1994ys}
C.~Hull, P.~Townsend, {Unity of superstring dualities}, Nucl. Phys. B 438
  (1995) 109--137.
\newblock \href {http://arxiv.org/abs/hep-th/9410167}
  {\path{arXiv:hep-th/9410167}}, \href
  {https://doi.org/10.1016/0550-3213(94)00559-W}
  {\path{doi:10.1016/0550-3213(94)00559-W}}.

\bibitem{Witten:1995ex}
E.~Witten, {String theory dynamics in various dimensions}, Nucl. Phys. B 443
  (1995) 85--126.
\newblock \href {http://arxiv.org/abs/hep-th/9503124}
  {\path{arXiv:hep-th/9503124}}, \href
  {https://doi.org/10.1016/0550-3213(95)00158-O}
  {\path{doi:10.1016/0550-3213(95)00158-O}}.

\bibitem{Izaurieta:2011fr}
F.~Izaurieta, E.~Rodriguez, {On eleven-dimensional Supergravity and
  Chern-Simons theory}, Nucl. Phys. B 855 (2012) 308--319.
\newblock \href {http://arxiv.org/abs/1103.2182} {\path{arXiv:1103.2182}},
  \href {https://doi.org/10.1016/j.nuclphysb.2011.10.012}
  {\path{doi:10.1016/j.nuclphysb.2011.10.012}}.

\bibitem{Jona-Lasinio:1964zvf}
G.~Jona-Lasinio, {Relativistic field theories with symmetry breaking
  solutions}, Nuovo Cim. 34 (1964) 1790--1795.
\newblock \href {https://doi.org/10.1007/BF02750573}
  {\path{doi:10.1007/BF02750573}}.

\bibitem{Adler:1969er}
S.~L. Adler, W.~A. Bardeen, {Absence of higher order corrections in the
  anomalous axial vector divergence equation}, Phys. Rev. 182 (1969)
  1517--1536.
\newblock \href {https://doi.org/10.1103/PhysRev.182.1517}
  {\path{doi:10.1103/PhysRev.182.1517}}.

\bibitem{Zee:2003mt}
A.~Zee, {Quantum field theory in a nutshell (Second Ed.)}, Princeton University
  Press, 2010.

\bibitem{Scharnhorst:1993bk}
K.~Scharnhorst, {Functional integral equation for the complete effective action
  in quantum field theory}, Int. J. Theor. Phys. 36~(2) (1997) 281--343.
\newblock \href {http://arxiv.org/abs/hep-th/9312137}
  {\path{arXiv:hep-th/9312137}}, \href {https://doi.org/10.1007/BF02435737}
  {\path{doi:10.1007/BF02435737}}.

\bibitem{Scharnhorst:2023spc}
K.~Scharnhorst, {On self-consistency in quantum field theory} (1 2023).
\newblock \href {http://arxiv.org/abs/2301.13275} {\path{arXiv:2301.13275}}.

\bibitem{Wilczek:1983cy}
F.~Wilczek, A.~Zee, {Linking Numbers, Spin, and Statistics of Solitons}, Phys.
  Rev. Lett. 51 (1983) 2250--2252.
\newblock \href {https://doi.org/10.1103/PhysRevLett.51.2250}
  {\path{doi:10.1103/PhysRevLett.51.2250}}.

\bibitem{Wu:1984kd}
Y.-S. Wu, A.~Zee, {Comments on the Hopf Lagrangian and Fractional Statistics of
  Solitons}, Phys. Lett. B 147 (1984) 325--329.
\newblock \href {https://doi.org/10.1016/0370-2693(84)90126-6}
  {\path{doi:10.1016/0370-2693(84)90126-6}}.

\bibitem{Nepomechie1984}
R.~Nepomechie, A.~Zee, Phase interaction between extended objects, in:
  I.~Batalin, C.~Isham, G.~Vilkovisky (Eds.), Quantum Field Theory and Quantum
  Statistics. Essays in Honour of the Sixtieth Birthday of E. S. Fradkin.,
  Hilger, 1987, pp. 467--472.

\bibitem{Wu:1988py}
Y.~S. Wu, A.~Zee, {Membranes, Higher Hopf Maps, and Phase Interactions}, Phys.
  Lett. B 207 (1988) 39--43.
\newblock \href {https://doi.org/10.1016/0370-2693(88)90882-9}
  {\path{doi:10.1016/0370-2693(88)90882-9}}.

\end{thebibliography}

\end{document}